# Magnetization dynamics in magnonic structures with different geometries: interfaces, notches and waveguides

M. Mansurova [1,2]  J. von der Haar[2], J. Panke[2], Jakob Walowski[1], Henning Ulrichs[2], M. Münzenberg[1]

1. Institut für Physik, Greifswald University, Felix-Hausdorff-Straße 6, 17489 Greifswald, Germany

2. I. Physikalisches Institut, University of Göttingen, Friedrich-Hund-Platz 1, 37077 Göttingen, Germany

*Abstract: The discovery of the ultrafast magnetization dynamics twenty years ago has led to a broad variety of experimental techniques to explore phenomena in magnetic materials with high temporal resolution. In the present article we present a study dealing with broadband excitation of spin-wave packets at different magnonic crystal continuous magnetic film interfaces. Similar to protected conducting states on the surfaces of topological band insulators, these interfaces exhibit surface spin-wave modes that propagate out of the crystal into the continuous film. The propagation distance depends on the direction of the applied magnetic field as well as the surface geometry of the crystal.*

**Introduction:**

One of the most important milestones that led to a major breakthrough in the modern field of magnetism was the observation of ultrafast magnetization switching in nickel by Bigot et al. [1]. This groundbreaking experiment not only led to a better understanding of the underlying fundamental processes that drive the magnetization reversal, but in addition to this, it opened a possibility to investigate a broad variety of magnetization phenomena using femtosecond ultrafast laser pulses to manipulate the spin-system dynamically. With that, it opened up new research directions and the knowledge gained over the years is leading to the first applications [2]. For example, it allowed developing the broadband excitation of spin-wave packets as was proposed in [3] by light, and allowed a further development of experiments that enable the investigation of spin waves in metallic thin films in the time domain with highest temporal resolution [4, 5, 6]. The first study on antidot magnonic crystals was presented in [7]. In such antidot magnetic crystal lattices, the magnetization inhomogeneities are very pronounced, making a fully analytical theoretical description of these systems impossible. The research field of magnonics has been established and investigated by other methods already earlier. The current state and the potentials of the research can be found in references [8, 9, 10].

In recent years the influence of topology on physical properties of particles and quasiparticles has awaken a broad interest. Topological band insulators can exhibit protected conducting states on their surfaces or edges [11] enabled by spin-orbit interaction and broken time-reversal symmetry. In the field of photonics, topologically protected states can be exploited to produce new states of light that bring useful and exciting properties into photonic metamaterials [12]. In the same manner, it has been proposed that a two-dimensional magnonic crystal can exhibit spin waves localized on the edge, as a result of magnonic spectrum only [13]. A topological magnonic crystal can exhibit protected chiral modes for magnetostatic spin waves as a result of dipolar [14] and Dzyaloshinskii-Moriya exchange interactions [15]. Those topologically protected spin waves propagating along one edge of a magnonic crystal (edge modes) are robust against lattice defects and boundary roughness, opening up the possibility of unidirectional, fault-tolerant spintronic devices. Transferring these concepts into spin-wave systems would open up a highly interesting field of research.

However, despite its attractive features, the experimental realization of such a topological magnonic crystal has not been provided to this date. Besides the difficulties such as the short spin wave coherence length and the careful design of the magnonic spectrum, so that the magnonic gap is opened with a nonzero sum of Chern numbers below the gap, there is a need to distinguish between the bulk and edge channels in a magnonic crystal. One of the most direct ways of doing so is the spin-wave detection by means of the spatially- and time-resolved magneto-optical Kerr effect measurements. This allows us to detect specific spin-wave states at the boundary of a magnonic crystal. Here we present the first step and demonstrate experiments at complex interface and notch structures at the boundary of magnonic crystals using the spatially resolved pump-probe technique of femtosecond laser-pulse generated spin waves.

**Sample preparation and experimental procedure:**

In this article, we connect the possibilities arising from ultrafast all-optical excitations in magnetic materials and different topologies within two dimensional magnetic antidot square lattices (magnonic crystals). We report a study on spatially-resolved magnetization dynamics measurements focusing on the magnetization precession frequencies observed at the edges of the crystals and their interfaces to continuous films. The samples are prepared as follows. A continuous 50 nm CoFeB film is magnetron-sputtered on a silicon substrate. Following this, a 2 nm Ru capping layer is e-beam evaporated on top. The antidots are produced by removing the material from the deposited thin film using focused ion beam (FIB). In all examined samples, the antidots have a 1 μm diameter and the separation between antidots is 3.5 μm. In order to characterize the bulk and edge channels in the magnonic crystal, four different structures are investigated. First, the interface between the magnonic crystal and the continuous film at different distances is examined. Second, we review a defect line in the magnonic crystal realized by several missing antidot rows. The last two discussed structures are magnonic crystal notches with 20 and 12 rows of antidots. For the measurements, the sample is situated in a saturating external magnetic field that is applied at 20° to the sample surface in order to tilt the internal field slightly out-of-plane. Magnetization dynamics is activated by a high-intensity ultrashort laser pulse with a duration of 40 fs and energy densities up to $40 mJ/cm^2$. The absorption of such a laser pulse rises the local temperature and triggers an effective anisotropy field pulse, which sets the spins inside the sample in motion [16]. The magnetization precession is then detected via the magneto-optical Kerr effect (MOKE) using a second, time-delayed (up to 1ns) and much weaker ($< 1 mJ/cm^2$) probe pulse. Both pump and probe beams have a Gaussian intensity profile and their full-width at half maximum (FWHM) at the sample surface amounts to 65 µm and 16 µm, respectively. For the analysis, the incoherent background is subtracted from the measured spectra and the time-resolved data is Fourier transformed into the frequency domain, enabling the extraction of the amplitude and frequency of the involved precession modes.

**Results and discussion:**

In Fig 1a, we show time-resolved magnetization dynamics excited and probed at different positions on the sample, moving onto the first structure. Both pump and probe beam stay overlapped by their centers, while being shifted across the interface between the magnonic crystal and the continuous CoFeB film for a distance of 100µm in steps of 10 µm. The scan direction is indicated by the blue arrow. The black lines show the measurements recorded on the magnonic crystal, while the red lines depict spectra measured in the region of the continuous film. The endpoints of the arrow shown in the right graph match the positions of the first and last measurement. For the lateral scan, the external magnetic field is kept constant at $\mu_0 H_{ext} = 130$ mT, being in-plane oriented at 45º with respect to the lattice vectors (see black arrow in Fig. 1a)) of the magnonic crystal and subsequently to the interface line between the magnonic crystal and the continuous film. In the data shown in Fig. 1 a), the incoherent background is subtracted for the dataset on each position. For the analysis, the resulting data is fast-Fourier transformed (FFT). The resulting Fourier power is shown in color code in Fig. 1b) (bottom). Comparing the detected spin-wave frequencies to those previously found on an identical magnonic crystals and continuous films [17], we can identify the uniform precession, the perpendicular standing spin wave (PSSW) and the additional Bloch or Damon-Eshbach (DE) like mode in the magnonic crystal. The data shows, that the uniform precession frequency on the magnonic crystal is slightly lower (0.5 GHz) than on the continuous film. This reduction can be explained in terms of a decreased internal magnetic field caused by the demagnetizing fields arising around the antidots. Please note that the DE mode is clearly detected up to 7.5 µm beyond the magnonic crystal interface in the continuous film due to the finite size of the probe beam.

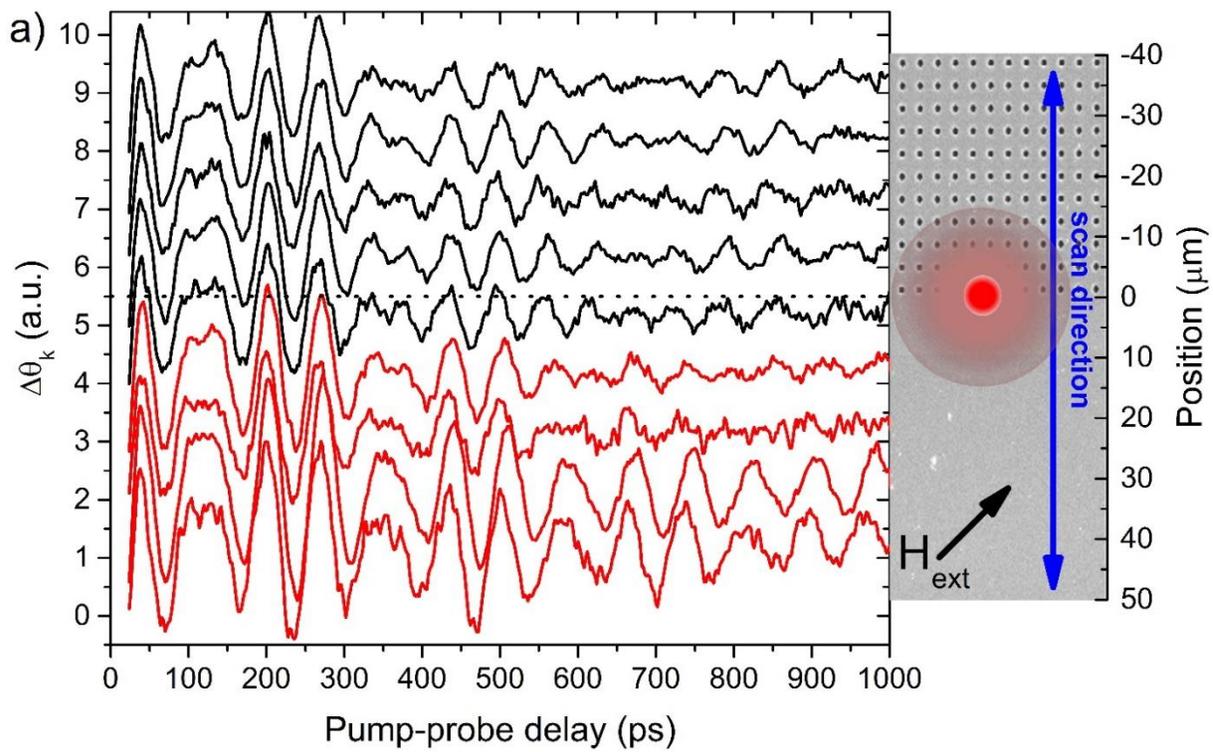

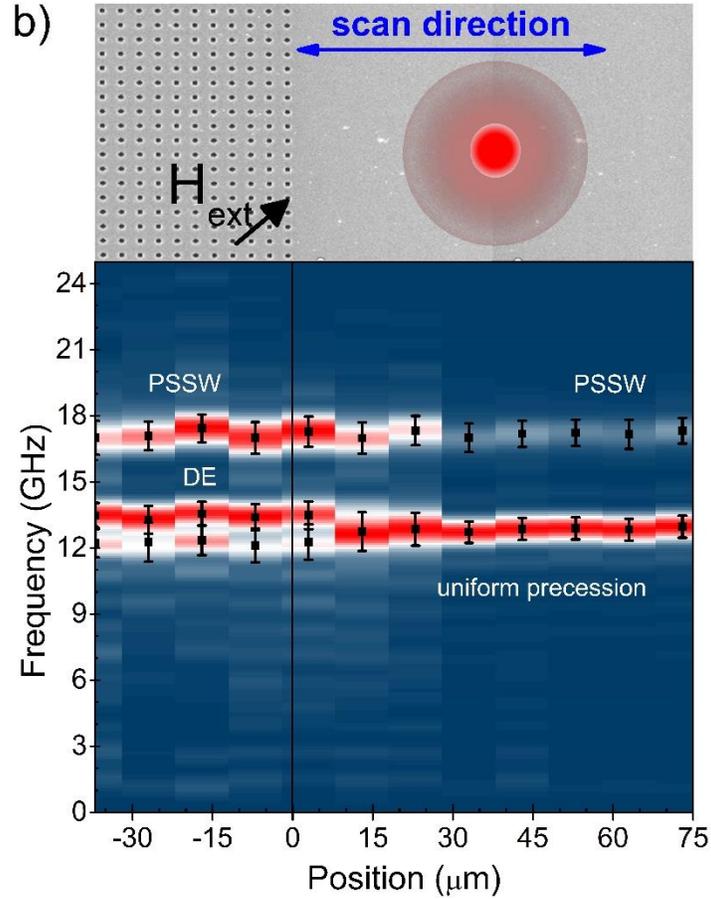

Fig. 1 a) Spatially-resolved magnetization dynamics across a magnonic crystal (black) and continuous film (red) interface. The right position graph depicts the investigated antidot interface. The end points of the spectra match the positions they were recorded according to the blue arrow in scan direction. The incoherent background is subtracted for each dataset. The Fourier transformed data for each position is shown in b). Each dataset is normalized to the highest peak of the power spectrum. On the continuous film uniform precession and PSSW precession frequencies are detected; on the magnonic crystal an additional DE mode, which can be tracked up to 10 μm beyond the antidot lattice, is found.

More complex patterns with different crystal surface geometries can be generated. In Fig. 2 we show the frequencies extracted from magnetization dynamics observed at different external magnetic fields in the middle point of a defect line, where five rows in a lattice structure are missing. This leaves a 20 μm total width of continuous film within the magnonic crystal. These are typical notch structures to observe topological surface states that are propagating 'around the corner' in photonic systems with a topological surface state in a magneto-photonic crystal [12]. The notch structure consists of a few missing lines of antidots, so called 'defect lines'. Here the focus is to investigate the states in the continuous film part in the middle of the notch forming by injection from the crystal boundary from two interfaces, which increases the complexity only slightly as compared to one interface. The graphs show color coded the amplitudes of the FFT power spectrum, the black squares, triangles and circles show the peak frequency values with error bars, and the white dotted lines are guides to the eye. In both graphs the insets show the defect line in the magnonic crystal where the measurements were performed, and the direction of the applied magnetic field. The measurements were carried out for two different orientations of the external magnetic field with respect to the lattice, at 45° and at 90°, as depicted in the insets of both graphs in Fig. 2. These data show that the DE mode can be detected beyond the magnonic crystal interface, propagating into the continuous film, within the spatial FWHM expansion of the probe beam. We can further show that the appearance of the DE mode can be controlled by the direction of the external magnetic field. We find no DE mode within the notch, when the field is applied at 45° angle to the long axis of the defect line, as can be seen in Fig. 2a. Whereas, when the field is perpendicular to the long axis of the notch, a clear additional spin wave mode is apparent at a slightly higher frequency than the uniform precession mode as shown in Fig. 2b. The frequency dependence of this mode on the applied field allows us to identify it as the Bloch mode like those previously found in the extended, bulk-like magnonic crystal. The associated wave vector is inversely proportional to the lattice constant $a$, $|\vec{k}| = \frac{\pi}{a}$ of the adjacent magnonic crystal which is the signature of a magnonic crystal Bloch mode. In the case, when the field is applied at 90° with respect to the lattice, the DE modes are generated at both ends of the notch and propagating towards each other, traveling a distance of 10µm till they meet. This distance coincides with the calculation made in [18], when assuming a higher group velocity outside of the magnonic crystal of $v_{gr} = 25 \frac{km}{s}$, the DE mode generated in the crystal is able to propagate distances in the order of 10µm to the middle of the notch. With this propagation distance out of the magnonic crystal for this kind of spin waves, the size of 20µm chosen for the notch width has the optimal size for forwarding the DE modes out of the crystal. Whereas in the case of the applied field aligned at 45° in respect to the lattice, the distance between two spin waves propagating towards each other increases up to ~ 30µm. That means they are damped before they meet. Furthermore, the DE modes generated in the configuration with the applied field aligned 45° in respect to the lattice, have a slightly smaller frequency, leading to a stronger decay within a distance of 10µm, and no transmission of that mode through the notch is found in this field configuration.

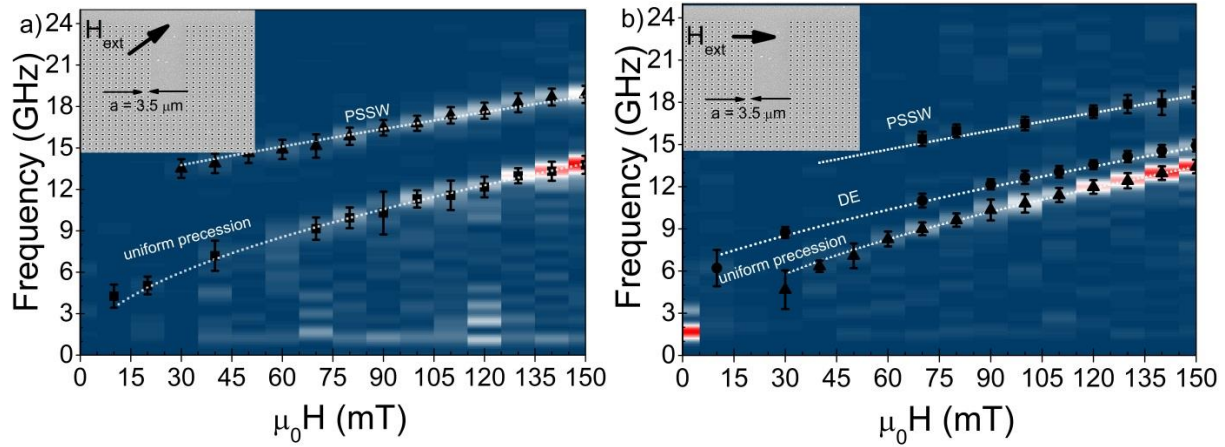

Fig. 2 FFT of magnetization dynamics on a defect line (five missing rows) in a magnonic crystal with an external magnetic field applied at a) 45º and b) 90º with respect to the defect line. Both, pump and probe beams are positioned and overlapped by their centers in the middle of the waveguide shown in the insets. The color coded maps show the intensity of the FFT power spectra, the black squares, triangles and circles show the extracted frequencies with the error bars. The white dotted lines are guides to the eye.

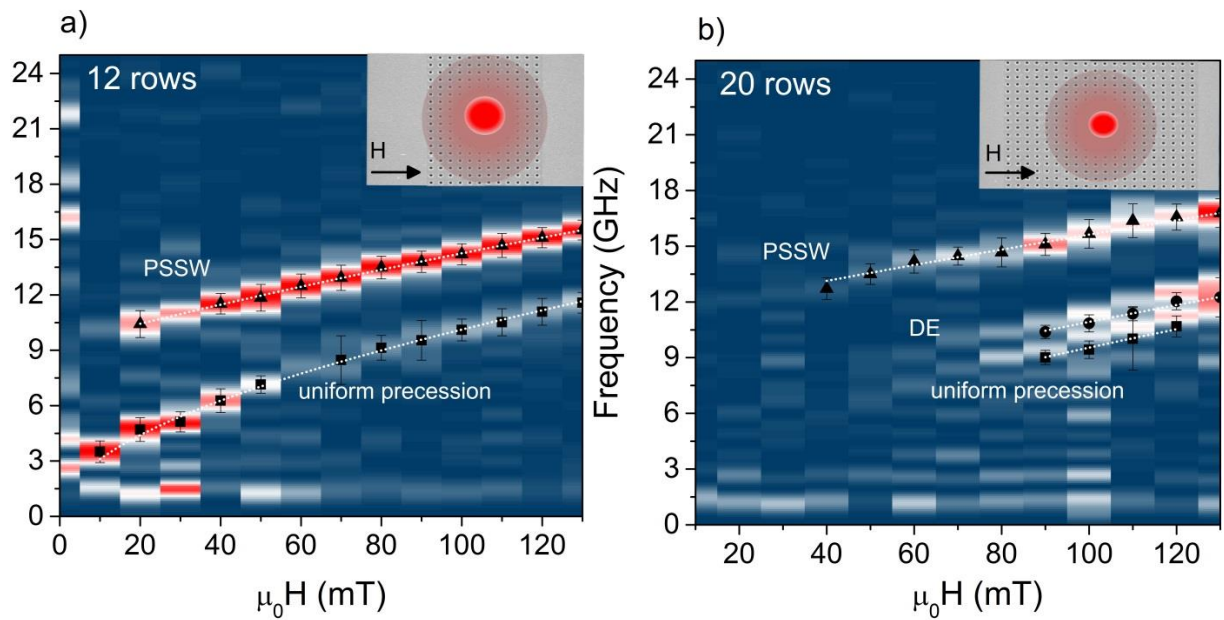

Fig 3: Magnetization dynamics on a magnonic crystal notch of a) 20 and b) 12 rows of antidots with the external magnetic field applied at 90º to the long axis of the notch. The insets show the position of pump and probe beams on the notch and the direction of the external magnetic field. The white dotted lines are guides to the eye.

Generally one can ask how large the magnonic crystal sizes have to be to form a magnonic crystal Bloch mode. Thus the last investigated structures are complementary stripes of magnonic antidot crystals that are tailored in size along one direction. The three investigated structures differ only by their total number of 6, 12 and 20 antidot rows. In Fig. 3 we present the data for 12 and 24 rows respectively. One observes the magnetization dynamics they exhibit is quite different. In this experiment, the pump and probe beams are positioned (overlapped by their centers) in the middle of the antidot rows. In the case of the 12 rows crystal, which has a width of 42 μm, the width of the area covered by the antidots is smaller than the pump beam FWHM diameter. But both antidot rows are larger than the FWHM diameter of the probe beam. The precession frequencies extracted from the dynamics measurements on the 12 rows crystal are shown in Fig. 3a). One finds only two spin wave modes: the uniform Kittel precession mode and the PSSW mode, i.e. a magnonic crystal Bloch mode is missing. In this respect the magnetization dynamics in this structure is identical to that of a continuous CoFeB film. Similar results were found for less than 12 rows, in an even smaller crystal of 6 rows. In contrast, the 20 rows crystal supports a Bloch-like DE precession mode, as can be seen in Fig. 3b). For the case of this antidot crystal based on a CoFeB film thus it seems that about 20 rows are needed to from the magnonic crystal Bloch mode. The measurements also show that the uniform Kittel precession mode is suppressed at small applied magnetic fields for the 20 rows crystal structure, whereas in the crystal structure with fewer rows, the uniform precession mode shows higher amplitudes at lower external magnetic fields.

**Conclusions:**

In summary, we have shown that magnetization dynamics in magnonic crystal structures can exhibit different modes that depend on the particular interface geometry. Although spatially-resolved measurements across a magnonic crystal/continuous film interface show Bloch-like modes outside the magnonic crystal, we find that on defect rows the excitation of the Bloch-like mode depends on the relative orientation of the external magnetic field with respect to the magnonic crystal lattice. This points to the importance of the symmetry of the excited Bloch-like state. Moreover, the size of the magnonic crystal also plays a determinant role in the formation of the Bloch mode. As shown in this work, only relatively large (compared to the size of the excitation spot) magnonic antidot crystals support the photo-induced formation of the Bloch-like modes.

**Acknowledgements:**

We thank the German Research Foundation (DFG) for funding through MU 1780/ 6-1 Photo-Magnonics, SPP 1538 SpinCaT and SFB 1073.